\documentclass{article}
\usepackage{graphicx} 
\usepackage{amsmath,amssymb}
\usepackage{siunitx}
\usepackage{authblk}
\usepackage{hyperref}
\newcommand{\dint}{{\rm d}}
\newcommand{\sartre}{Sar{\it t}re}
\newcommand{\xpom}{x_{I\!\!P}}

\title{Efficient calculation of exclusive diffractive cross sections at the EIC and LHeC with the \sartre~event generator}
\author{Tobias Toll, Dipan Ghosh, Abhinav Srivastav}
\affil{Department of Physics, Indian Institute of Technology Delhi, Hauz Khas, New Delhi 110 016,
India \\
Email: \href{mailto:tobiastoll@iitd.ac.in}{tobiastoll@iitd.ac.in} }

\begin{document}

\maketitle
\begin{abstract}
    We present a new version of the \sartre~ event generator for exclusive diffraction at small $\xpom$ in the colour dipole model, for $ep$ and $e$A scattering at the EIC and the LHeC as well as ultra-peripheral $pp$, $p$A, and AA collisions at RHIC and the LHC. \sartre~stores the first and second moment of the interaction amplitudes in lookup tables which are then used for efficient event generation. There are many possible combinations of processes in \sartre, with different initial state nuclear targets and different final state vector mesons or real photons. We also want to implement different versions of the dipole model: with and without non-linear saturation effects, with and without nucleon hotspot substructure. A long-standing bottleneck for simulating all possible exclusive processes has been the production of lookup tables, which take a few CPU-year for each combination, necessitating the use of computing farms. The calculation also involves integrations of rapidly fluctuating integrands, which may cause numerical glitches which takes much effort to smoothen out. In this paper we present a solution to these issues, by presenting a new numerical calculation which improves the efficiency in the table production by 3-4 orders of magnitudes. This enables us to produce lookup tables for any process that we may be interested in, in a few hours. The new calculation does also not exhibit numerical glitches. We provide novel predictions for the EIC and the LHeC using the new version of \sartre. 
\end{abstract}
\section{Introduction}
Exclusive Diffraction is seen as a key measurement for future Deeply Inelastic Scattering (DIS) experiments, such as the Electron-Ion Collider (EIC) ~\cite{Accardi:2012qut,AbdulKhalek:2021gbh} and the Large Hadron-electron Collider (LHeC)~\cite{LHeC:2020van}. These processes constitute excellent probes, both for non-linear evolution of the gluon structure described by gluon saturation, as well as for spatial gluon structure and substructure. \sartre~ \cite{Toll:2013gda,Toll:2012mb} is an event generator for these processes at small $x$ for $ep$ and $e$A scattering, as well as for ultra-peripheral $pp$, $p$A, and AA scattering at the LHC and RHIC experiments~\cite{Sambasivam:2019gdd, Nattoja:2025hze}. It has been used extensively in the preparations for the EIC. \sartre~ calculates cross sections using the dipole model framework, which is valid for $x<0.01$. 

There are two possible processes in  diffractive scattering, either the virtual photon interacts coherently with the entire target (proton or nucleus), or it interacts coherently only with a subregion of the target. In the latter case, called incoherent diffraction, the nucleus gets excited and subsequently de-excites, mostly by emitting neutrons or breaking up. According to the Good-Walker picture~\cite{Good:1960ba}, the coherent scattering cross section depends on the \emph{average} of the dipole amplitude with respect to the inital state gluon distribution, while the incoherent cross section depends on the \emph{variance} of the amplitude. In order to generate these events one therefore needs to calculate the first and the second moments of the amplitude with respect to the configuration of nucleons in the nucleus, as well as hotspots within the nucleons.

Calculating these moments has been a long standing bottle-neck for incoherent diffraction. We have earlier shown \cite{Toll:2012mb} that for most parts of phase-space, one needs to average over $\sim 500$ nucleon configurations. This constitutes evaluating two four-dimensional integrals for each phase-space point 500 times, which is prohibitive for efficient event generation. 

In \sartre~we have solved this by creating lookup tables for the first and second moments of the amplitude. This then ensures very efficient event generation. However, these lookup tables are incredibly cumbersome to produce, taking CPU-years for each combination of initial nuclear species and final state vector meson, necessitating heavy parallelization on computing farms. What further complicates the issue is that the integrands contain quickly fluctuating functions in some parts of phase-space, which makes the numerical intgration routines slow to converge. This also causes numerical glitches and spikes in some bins. To smoothen the lookup tables post production also takes a significant effort.

These issues were partially solved in a recent publication~\cite{Singh:2023yvj} where we employed neural network techniques. We showed that it is enough to create a lookup table containing only 10\% of the bins, and train a fairly small neural network on this. The neural network model is then used to fill out the rest of bins in the tables. The neural network is also excellent at smoothening out the resulting tables, such that numerical glitches and spikes does not affect the final event generation.

In this paper we present a significant update on the numerical calculation. The largest improvement comes from taking advantage of the fact that the integral constitutes a Fourier transform, which we turn into a discreet Fourier transform. This means that we can calculate the integral only on the phase-space points to be stored in the tables, and the numerical integration only need to take into account it's reciprocal points. 
We also take full advantage of the structure of the calculation in order to make it as fast as possible, by precomputing all parts of the calculation that are used more than once in each phase-space point. We show that we reach speed-ups in the calculations of the amplitude modes of 3-4 orders of magnitude. This corresponds to a reduction from CPU-years to less than a CPU-day. Combining it with the parallelization available on most computers today, we have now moved the \sartre~ amplitude lookup table production from computing farms to users' laptops.

This speedup enables \sartre~to simulate all combinations of intial state nuclear targets and final state vector mesons or DVCS, with and without non-linear saturation physics, and with and without nucleon hotspot substructure. In the earlier version, one had to make choices for which processes to produce lookup tables for event-generation. In this version, these choices are no longer needed. 

The paper is organized as follows. In the next session we describe the underlying dipole formulation and describe the structure of the calculation. In section \ref{sec:optimization} we describe how we organize the numerical implementation in an efficient way. In section \ref{sec:results} we present a comparison with the previous version of \sartre~to show that the new version produces the same results. We also present a study of 3-4 order of magnitude improved efficiency. Then, in section \ref{sec:predictions} we present completely new physics predictions for the EIC and the LHeC.

\section{Structure of the calculation}
The total cross section is given by the second moment of the dipole interaction amplitude:
\begin{eqnarray}
    \frac{\dint\sigma_{T, L}}{\dint t}=\frac{1}{16\pi}\left<\left|\mathcal{A}_{T, L}(x, Q^2, t, \Omega)\right|^2\right>
    \label{eq:dsigmadt}
\end{eqnarray}
The bracket on the right-hand side is the object that is tabulated in \sartre, together with the first moment of the amplitude.
Here,
\begin{eqnarray}
    \mathcal{A}_{T,L}(x, Q^2, t, \Omega)&=&i\int {\rm  d}r\int{\rm d}z\left(\psi^*_V\psi\right)_{T, L}(r, z)2\pi J_0([0.5-z]r\Delta)\cdot \nonumber \\
    &\cdot& \int{\rm d}^2\vec b\frac{{\rm d}\sigma}{{\rm d}^2b}(x, r, \vec b, \Omega)e^{-i\vec b\cdot \vec \Delta}
\end{eqnarray}
where $r$ is the dipole transverse size, $z$ is the photon's momentum fraction carried by the quark in the dipole, $\left(\psi^*_V\psi\right)$ is the photon-vector meson wave overlap, and $\frac{{\rm d}\sigma}{{\rm d}^2b}$ is the so called dipole cross section, which is twice the dipole amplitude. 
In \sartre, two versions of the dipole cross section are implemented, IPsat, and IPnosat~\cite{Kowalski:2003hm,Kowalski:2006hc}. 

\subsection{Proton targets}
For a proton target (without substructure) the dipole cross sections are given by:
\begin{eqnarray}
    \frac{{\rm d}\sigma^{\rm IPsat}_p}{{\rm d}^2b}&=&2\left[1-e^{-\Omega(\xpom, r, b)/2}\right] 
    \\
    \frac{{\rm d}\sigma^{\rm IPnosat}_p}{{\rm d}^2b}&=&\Omega(\xpom, r, b)
\end{eqnarray}
here, the opacity $\Omega(\xpom, r, b)=\pi^2r^2/N_C\alpha_S(\mu^2)\xpom g(\xpom,\mu^2)T(b)$, where the factorization and renormalization scales are both taken at $\mu^2=C/r^2+\mu_0^2$. The thickness function $T(b)$ is taken to be a Gaussian for the proton. The IPsat model clearly saturates at large $r$, $xg$, and $T$, while the IPnosat model is a linearized version of IPsat. 
Currently, there are three parametrization of these models corresponding to different fits to HERA data, namely KMW~\cite{Kowalski:2006hc}, HMPZ ~\cite{Mantysaari:2018nng}, and STU~\cite{Sambasivam:2019gdd}. The latter two are fitted to the combined H1 and ZEUS inclusive reduced cross sections~\cite{Abramowicz:2015mha} as well as $F_2^c$~\cite{H1:2018flt} measurements. It should be noted that the STU parametrization contains a phenomenological Gaussian damping of large dipoles in the dipole amplitude to suppress unphysically large dipoles. In this paper we show results using the STU parametrization.

\subsection{Nuclear Targets}
For nuclear targets, \sartre~ implements the independent scattering approximation, where 
\begin{eqnarray}
    1-\frac12 \frac{{\rm d}\sigma_A}{{\rm d}^2b}(\xpom, r, \vec b)=\prod_{i=1}^A\left(1-\frac12 \frac{{\rm d}\sigma_p}{{\rm d}^2b}(\xpom, r, |\vec b-\vec b_i|)\right)
\end{eqnarray}
where $A$ is the nucleus mass number, and the nucleon positions $b_i$ are distributed according to a Woods-Saxon distribution. For the IPsat and IPnosat models, this manifests in a modified thickness function:
\begin{eqnarray}
    T_A(\vec b)=\sum_{i=1}^A T_p(|\vec b-\vec b_i|)
    \label{eq:nuclearthickness}
\end{eqnarray}
To get the cross section in eq.\eqref{eq:dsigmadt}, one has to average the absolute amplitude square over all possible configurations of nucleons inside the nucleus. In practice it has been shown that $\sim 500$ configuration gives a reasonable convergence up to moderate values of $|t|$ ~\cite{Toll:2012mb}.

However, according to the Good-Walker mechanism~\cite{Good:1960ba}, there are two components to the total cross section in eq.\eqref{eq:dsigmadt}, the coherent and incoherent cross sections, where:
\begin{eqnarray}
    \left<|\mathcal{A}|^2\right>=\left<\mathcal{A}\right>^2 + \left[\left<|\mathcal{A}|^2\right> - \left<\mathcal{A}\right>^2\right]
\end{eqnarray}
Here, the first term on the right hand side corresponds to coherent interactions where the target stays intact, while the variance term in the square brackets corresponds to the incoherent cross section where the target gets excited in the interaction and subsequently decays. In order to capture this physics with \sartre ~we therefore need to calculate both the first and second moments of the amplitude and store them in lookup tables.

A computational challenge comes from the fact that for already moderately large values of $|t|$, the incoherent cross section (variance) becomes orders of magnitude larger than the coherent cross section (mean), which tells us that the convergence of the first moment becomes very slow. In this region we instead use the optical approximation:
\begin{eqnarray}
    \left<\frac{{\rm d}\sigma_A}{{\rm d}^2b}\right>_\Omega 
    =2\left[1-\left(1-\frac{T_A(b)}{2}\sigma_p\right)^A\right] \label{eq:optical}
\end{eqnarray}
where $T_A(b)$ is the two-dimensional projection of the Woods-Saxon distribution, and $\sigma_p=\int\dint^2\vec b  ~{\rm d}\sigma_p/{\rm d}^2b$.

\subsection{Nucleon Substructure}
We can use the independent scattering approximation to also include nucleon hotspot substructure ~\cite{Mantysaari:2016jaz,Mantysaari:2016ykx} by modifying the thickness function in the following way:
\begin{eqnarray}
    T_p(\vec b)=\frac{1}{N_{hs}}\sum_{j=1}^{N_{hs}}T_{hs}(|\vec b_j-\vec b|)
    \label{eq:hotspotthickness}
\end{eqnarray}
where the hotspot positions $b_j$ are usually distributed according to a Gaussian. We take $N_{hs}=3$ and the parameters from ~\cite{Kumar:2024kns}.

\subsection{Phenomenological Corrections to the Dipole Cross section}
The derivation of the amplitude relies on the approximation that the $S$-matrix is purely imaginary. However, one may take into account the real part of the $S$-matrix by multiplying the cross section by a factor $(1+\beta^2)$, where $\beta$ is the real-to-imaginary ratio of the amplitude given by \cite{Kowalski:2006hc}:
\begin{eqnarray}
    \beta=\tan\left(\frac\pi2\lambda\right), 
    ~\lambda=\frac{\partial\log\mathcal{A}_{T,L}^{\gamma^*p}}{\partial\log x}
    \label{eq:real}
\end{eqnarray}

The two gluons in the interaction may carry different momentum fractions. This skewedness correction can be accounted for by multiplying the cross sections by a factor $R_g$ given by:
\begin{eqnarray}
    R_g(\lambda)=\frac{2^{2\lambda+3}}{\sqrt\pi}\frac{\Gamma(\lambda+5/2)}{\Gamma(\lambda+4)},~~\lambda=\frac{\partial\log xg(x,\mu^2)}{\partial\log x}
    \label{eq:skew}
\end{eqnarray}
This skewedness correction is strictly only true for non-linear versions of the dipole model.

\section{Optimising the computation}
\label{sec:optimization}
In \sartre, we tabulate the first two moments of the amplitude, averaged over 500 configurations, in bins of $Q^2$, $W^2$, and $t$. These 3D tables typically contain 10,000-30,000 bins. For each bin we need to evaluate the four-dimensional integral, averaged over 500 configurations, for two photon polarizations. These integrations contain functions $J_0([0.5-z]r\Delta)$ and $\exp(-i|\vec b||\vec\Delta|\cos(\phi_{b\Delta}))$, which are rapidly oscillating at large $t$.

In previous versions of \sartre, this process can sometimes take weeks of dedicated production time on computing farms, making it prohibitive to create a vast number of tables, as well as to implement changes in the models. In this section we will show how this process can be sped up by 3-4 orders of magnitude.

\subsection{Utilizing the Fourier transform structure of the calculation}
The main idea is to utilize the properties of the Fourier transform (FT) to our advantage. We may write the amplitude in the following form:
\begin{eqnarray}
    \mathcal{A}_{T, L}(\xpom, Q^2, t, \Omega)=i\int\dint r K_{T, L}(r)\tilde F(r, \vec\Delta) 
\end{eqnarray}
where
\begin{eqnarray}
    \tilde F(r, \vec\Delta)=\int{\rm d}^2\vec b\frac{{\rm d}\sigma}{{\rm d}^2b}(x, r, \vec b, \Omega)e^{-i\vec b\cdot \vec \Delta}
\end{eqnarray}
and 
\begin{eqnarray}
    K_{T, L}(r)=\int\dint z\left(\psi^*_V\psi\right)_{T, L}(r, z)2\pi J_0([0.5-z]r\Delta)
\end{eqnarray}
The impact parameter integral in $\mathcal{A}_{T, L}$ is a Fourier transform from $\vec b$ to $\vec\Delta$. One way to utilize this is to form a Discreet Fourier Transform (DFT), or a Fast Fourier Transform (FFT). A key property of the FFT is that it creates a grid in $b_x$ and $b_y$ on which it evaluates the function and gives the Fourier transform as a reciprocal grid in $\vec\Delta$-space. This means that the algorithm creates a whole array of $t$-values as $t=-(\Delta_x^2+\Delta_y^2)$. If we choose our FFT-parameters carefully, we can then directly save the entire grid as our t-values in the table. By utilizing this property of the FFT-algorithm we therefore reduce the problem from calculating our results in $N_{Q^2}\times N_{W^2} \times N_t\sim20\times 20\times 40$ 3D bins in $(Q^2, W^2, t)$ to a 2D problem where we make a calculation only in $(Q^2, W^2)$ and get all the $t$-bins directly from the FFT evaluation. This by itself speeds up the table generation by a factor of $N_t$, where typically $N_t\sim 30-40$. However, we may also use the projection-slice theorem to simplify the routine, since $\tilde F(r, \vec\Delta)=\tilde F(r, |\vec \Delta|)$, by taking a slice in $\Delta_x$:
\begin{eqnarray}
    \tilde F(r, |\vec\Delta|)=\tilde F(r, \Delta_x)\big|_{\Delta_y=0}=\int\dint b_x\left[\int\dint b_y \frac{{\rm d}\sigma}{{\rm d}^2b}(b_x, b_y)\right]e^{-2\pi i\Delta_xb_x}
\end{eqnarray}
Then the routine reduces to a 1D FT. 
Since we are calculating the Fourier transform on a grid in $\vec b$-space, we need $N_b^2$ function calls for the inner sum, and $N_bN_t/2$ calls for the DFT step, which becomes $N_b\log_2(N_b)$ calls for an FFT.

For the parts of the cross-section that have azimuthal symmetry in the impact parameter, we may instead use a 1D Hankel transform.
Here, the integral over impact parameter becomes:
\begin{eqnarray}
    H(r,\Delta)=2\pi\int b\dint b \frac{{\rm d}\sigma}{{\rm d}^2b}(x, r, b)J_0(b\Delta).
    \label{eq:Hankel}
\end{eqnarray}

\subsection{Implementation}
The amplitude for a single configuration $\Omega$ becomes:
\begin{eqnarray}
    \mathcal{A}_{T, L}=\frac i2\sum_{i=1}^{N_r}\omega_{r_i} r_i K_{T, L}[n](r_i)\tilde F_n(r_i)
\end{eqnarray}
where we use Gauss-Legendre (G-L) 1D integration and $\omega_{r_i}$ are the G-L weights. 

Similarly, we use G-L to compute the kernels $K_{T,L}$:
\begin{eqnarray}
    K_{T, L}[n](r_i)=\sum_{j=1}^{N_z}\omega_{z_j}(\psi^*_V\psi)_{T, L}(r, z_j, Q^2)J_0\left((1/2-z_j)r\Delta_n\right)
\end{eqnarray}
Here, the Bessel function argument ranges from 0 at $z=1/2$ to $r_{\rm max}\Delta_{N_b/2}/2=\pi N_b/4$. If we require at least two nodes per oscillation in the G-L integration, this gives $N_z=N_b/2$. We set it adaptively to $N_z=\max(32, N_b/2)$. 
The kernels $K_{T, L}$ are common for each nuclear configuration and can therefore be calculated once for each kinematic point in $(Q^2, W^2)$. This means that we are left with a 1-dimensional integration over $z$ for each phase-space point and configuration. 

The opacity in the dipole cross section can be separated into a b-dependent and a b-independent part:
\begin{eqnarray}
    \Omega(\xpom, r, \vec b)=\rho(\xpom, r)T(\vec b)
\end{eqnarray}
Here, again, all nuclear configurations use the same $\rho(\xpom, r)$ which can be calculated once at each kinematic point. This assumes that $\xpom$ is independent of $t$. There is, however, a small $t$-dependence in $\rho(\xpom, r)$, since $\xpom=\xpom(t, Q^2, W^2)$, leading to a slow $t$ dependence of $xg(x,\mu^2)$. We describe below how this $t$-dependence in $\rho$ is treated.

The Fourier transform step looks as follows:
\begin{eqnarray}
    \tilde F(r, \Delta_{x(n)})\big|_{\Delta_y=0}=\left[\sum_{j=0}^{N_b-1}\sum_{k=0}^{N_b-1}\frac{{\rm d}\sigma_{jk}}{{\rm d}^2b}(r)\cos\left(\frac{2\pi nj}{N}\right)\right](\delta b)^2(-1)^n \nonumber \\
    -i\left[\sum_{j=0}^{N_b-1}\sum_{k=0}^{N_b-1}\frac{{\rm d}\sigma_{jk}}{{\rm d}^2b}(r)\sin\left(\frac{2\pi nj}{N}\right)\right](\delta b)^2(-1)^n
\end{eqnarray}
Using the symmetries of the sine and cosine functions, this yields $N_b\times N_t/2$ functions calls. 

It should be noted that this is a discreet Fourier transform (DFT). If we turn it into an FFT, the number of function calls reduces to $\mathcal{O}(N_b\log(N_b))$, while the DFT scales as $N_t\times N_b$, where $N_t$ is the number of $t$-bins we are calling the algorithm with. The user has a choice between using the DFT version which enables arbitrary $t$-values for the table, or the FFT version which fixes the $t$ values at the reciprocal $\Delta$-grid. The reciprocal grid is given by, $\Delta_{x(n)}=n\delta\Delta$, where $\delta\Delta=2\pi/(N\delta b)=\pi/b_{\rm max}$, and $n\in[1,N_b/2]$. A further point to note is that the actual information of the Fourier transform sits at the reciprocal grid points, so using arbitrary $\Delta$-values constitutes an interpolation. Given these considerations, we have implemented both the DFT and the FFT calculations as options in \sartre, where in the former case, the $t$-values are determined by the user. 

We precompute the $b$-dependence for each configuration on a $N_b\times N_b$ dimensional grid:
\begin{eqnarray}
    T_{jk}=T(b_{x(j)}, b_{y(k)})
\end{eqnarray}
where
\begin{eqnarray}
    b_{x(j)}=(j-N_b/2)\delta b, ~
    b_{y(k)}=(k-N_b/2)\delta b,~{\rm and}~\delta b=2b_{\rm max}/N_b \nonumber \\
\end{eqnarray}
using eqs. \eqref{eq:nuclearthickness} and \eqref{eq:hotspotthickness} for the functional form of the thickness. 

We also pre-compute $\rho(\xpom, r)$ on a grid in $\xpom(t)$ and $r$, which is common for all nuclear configurations. We observe that the true value of $\xpom$ at a given $\Delta$ grid point $n$ is given by:
\begin{eqnarray}
    \xpom(n)=\xpom(0)+\frac{n^2(\delta\Delta)^2}{W^2+Q^2-m_p^2}
\end{eqnarray}
where $m_p$ is the proton mass, and $\xpom(0)=\xpom(t=0)$. The approximation $\xpom=\xpom(t=0)$ is good for small $\xpom$ and small $|t|$. We implement the correction to this by tabulating $\rho(\xpom, r)$ for $n_l$ values in $\xpom$, equally spaced in $\log(\xpom)$ and then doing a log-linear interpolation on this table. We choose $n_l$ dynamically by:
\begin{eqnarray}
L=\log(\xpom(N_b/2)/\xpom(0)), ~~~
    n_l=\begin{cases}1~~:~~ L<0.05 \\
    2~~:~~0.05\leq L <0.14 \\
    4~~:~~0.15\leq L <0.40 \\
    8~~:~~0.40 \leq L
    \end{cases}
    \label{eq:slab}
\end{eqnarray}
This will increase the computational time by a factor of 8 for some kinematic points.

For the Hankel transform in eq.\eqref{eq:Hankel} we again use the Gauss-Legendre Quadrature method. This is used for calculating the amplitude with protons without substructure, which is also used for calculating the real-part and skewedness corrections in eqs.\eqref{eq:real} and \eqref{eq:skew}, as well as the optical approximation for coherent diffraction in eq.~\eqref{eq:optical}.

\section{Results}
\label{sec:results}
\subsection{Comparisons with Cuhre}
The previous version of \sartre~used the Cuhre algorithm in the Cuba library \cite{Hahn:2005pf} for the 4D integral over $z$, $r$, and $\vec b$. Cuhre is a adaptive integration routine that employs cubature rules of varying polynomial degrees to iteratively subdivide the integration domain. It works by splitting the integration domain into subregions and then calculates the integral on each subregion. The subregion with the largest error is further divided, until all subregions have reached the desired precision. It estimates the error by the difference between the evaluation of the integral using a polynomial of order $p$ and and using a polynomial of order $p-1$.

Since the integrand contains fluctuating functions $J_0((1/2-z)r\Delta)$ and $\exp(i\Delta b\cos(\phi_{\Delta b}))$, this can force Cuhre to make many subdivisions on some parts of phase space. In practice one needs to find a good compromise between speed and precision for integration routines, and the default setting in \sartre~involved demanding a relative error in each integral of $1\%$, and capping the integration at $10^9$ function calls. Cuhre also provide its estimated error to its output.

\begin{table}
\begin{center}
\textbf{$Q^2=2~\mathrm{GeV}^2$, $W=40~\mathrm{GeV}$}, time for 64 FFT bins: 26 ms

\vspace{0.5em}

\resizebox{\columnwidth}{!}{%
\begin{tabular}{|c||c  |c | c|c|| c | c| c|c|}
\hline
 & \multicolumn{4}{c||}{$|A_T|^2$}  
      & \multicolumn{3}{c|}{$|A_L|^2$}  &  \\
\cline{2-5} \cline{6-9}
$|t|$ & Cuhre & err (\%) & FFT & ratio  
 & Cuhre & err (\%) & FFT & ratio \\
\hline
0.006 & $1.004\cdot10^{-1}$ & 0.05 & $1.004\cdot10^{-1}$ & 1.000
      & $1.249\cdot10^{-2}$ & 0.04 & $1.248\cdot10^{-2}$ & 0.999 \\
0.022 & $2.145\cdot10^{-4}$ & 1.20 & $2.128\cdot10^{-4}$ & 0.992
      & $2.29\cdot10^{-5}$  & 0.91 & $2.264\cdot10^{-5}$ & 0.989 \\
0.140 & $4.348\cdot10^{-5}$ & 1.71 & $4.422\cdot10^{-5}$ & 1.017
      & $4.685\cdot10^{-6}$ & 1.76 & $4.788\cdot10^{-6}$ & 1.022 \\
1.099 & $3.206\cdot10^{-4}$ & 1.45 & $3.251\cdot10^{-4}$ & 1.014
      & $4.103\cdot10^{-5}$ & 1.04 & $4.148\cdot10^{-5}$ & 1.011 \\
2.242 & $1.82\cdot10^{-4}$  & 1.89 & $1.805\cdot10^{-4}$ & 0.992
      & $2.356\cdot10^{-5}$ & 1.73 & $2.36\cdot10^{-5}$  & 1.002 \\
\hline
\end{tabular}
}

\vspace{1em}

\textbf{$Q^2=10~\mathrm{GeV}^2$, $W=75~\mathrm{GeV}$}, time for 64 FFT bins: 14 ms

\vspace{0.5em}

\resizebox{\columnwidth}{!}{%
\begin{tabular}{|c||c | c | c|c|| c | c| c|c|}
\hline
 & \multicolumn{4}{c||}{$|A_T|^2$}  
      & \multicolumn{3}{c|}{$|A_L|^2$}  &  \\
\cline{2-5} \cline{6-9}
$|t|$ & Cuhre & err (\%) & FFT & ratio  
 & Cuhre & err (\%) & FFT & ratio \\
\hline
0.006 & $3.398\cdot10^{-2}$ & 0.08 & $3.39\cdot10^{-2}$ & 0.998
      & $2.056\cdot10^{-2}$ & 0.04 & $2.027\cdot10^{-2}$ & 0.986 \\
0.022 & $6.702\cdot10^{-5}$ & 1.91 & $6.366\cdot10^{-5}$ & 0.950
      & $3.397\cdot10^{-5}$ & 1.18 & $3.26\cdot10^{-5}$  & 0.960 \\
0.140 & $1.316\cdot10^{-5}$ & 1.75 & $1.343\cdot10^{-5}$ & 1.021
      & $7.037\cdot10^{-6}$ & 1.80 & $7.035\cdot10^{-6}$ & 1.000 \\
1.099 & $1.142\cdot10^{-4}$ & 1.59 & $1.136\cdot10^{-4}$ & 0.995
      & $6.968\cdot10^{-5}$ & 1.10 & $7.05\cdot10^{-5}$  & 1.012\\
2.242 & $6.525\cdot10^{-5}$ & 1.88 & $6.51\cdot10^{-5}$  & 0.998
      & $4.101\cdot10^{-5}$ & 1.48 & $4.132\cdot10^{-5}$ & 1.008 \\
\hline
\end{tabular}
}

\vspace{1em}

\textbf{$Q^2=50~\mathrm{GeV}^2$, $W=140~\mathrm{GeV}$}, time for 64 FFT bins: 8 ms

\vspace{0.5em}

\resizebox{\columnwidth}{!}{%
\begin{tabular}{|c||c  | c | c|c|| c | c| c|c|}
\hline
 & \multicolumn{4}{c||}{$|A_T|^2$}  
      & \multicolumn{3}{c|}{$|A_L|^2$}  &  \\
\cline{2-5} \cline{6-9}
$|t|$ & Cuhre & err (\%) & FFT & ratio  
 & Cuhre & err (\%) & FFT & ratio \\
\hline
0.006 & $1.243\cdot10^{-3}$ & 0.05 & $1.241\cdot10^{-3}$ & 0.998
      & $3.492\cdot10^{-3}$ & 0.05 & $3.493\cdot10^{-3}$ & 1.000 \\
0.022 & $1.38\cdot10^{-6}$  & 1.72 & $1.328\cdot10^{-6}$ & 0.962
      & $3.316\cdot10^{-6}$ & 1.89 & $3.286\cdot10^{-6}$ & 0.991 \\
0.140 & $3.079\cdot10^{-7}$ & 1.93 & $3.055\cdot10^{-7}$ & 0.992
      & $7.665\cdot10^{-7}$ & 1.96 & $7.808\cdot10^{-7}$ & 1.019 \\
1.099 & $4.438\cdot10^{-6}$ & 1.55 & $4.436\cdot10^{-6}$ & 1.000
      & $1.261\cdot10^{-5}$ & 1.53 & $1.267\cdot10^{-5}$ & 1.005 \\
2.242   &     $2.701\cdot10^{-6}$ & 1.85 &    $2.718\cdot 10^{-6}$ &  1.007 &  $7.782\cdot10^{-06}$ &  1.85 &  $7.882\cdot10^{-06}$ &  1.013 \\
\hline
\end{tabular}
}
\end{center}
\caption{Comparison between the two integration methods Cuhre and the DFT and FFT methods on a single configuration. $|t|$ is in (GeV)$^2$, $A_{T, L}$ is in fm$^2$.}
\label{tab:table1}
\end{table}

\begin{table}
\begin{center}
\resizebox{\columnwidth}{!}{%
\begin{tabular}{|c||c  |c | c|c|}
    \hline
    \multicolumn{5}{|c|}{Timings} \\\hline
     & $Q^2=2~{\rm GeV}^2$& $Q^2=10~{\rm GeV}^2$ & $Q^2=50~{\rm GeV}^2$ & Speedup \\
    Integration Routine & $W=40~{\rm GeV}$ & $W=75~{\rm GeV}$ & $W=140~{\rm GeV}$ & Factor\\ \hline
    Cuhre $N_t=5$ & 36s & 43s & 103s & 1\\ \hline
    DFT, $N_b=128, N_t=5$ & 14ms & 7ms & 4ms  & 7,100\\ 
    DFT, $N_b=256, N_t=5$ & 49ms & 25ms & 13ms & 2,000\\
    \hline
    FFT, $N_b=128$, $N_t=64$ & 29ms & 15ms & 8ms & 3,500\\
    FFT, $N_b=256$, $N_t=128$ & 100ms & 100ms & 51ms & 720\\ \hline
\end{tabular}
} 
\end{center}
\caption{An overview of the calculation times for the bins shown in table \ref{tab:table1}. The FFT timings are for all 64 and 128 points in $t$ respectively.}
\label{tab:times}
\end{table}

In table \ref{tab:table1} we present a comparison of the results between the previous version using the Cuhre calculation, and the new DFT/FFT implementation, for three kinematic points in $Q^2$ and $W^2$. The FFT implementation generates the amplitude for $N_b/2$ points in $t$ simultaneously, while Cuhre calculates the entire integral for each point in $Q^2$, $W^2$, and $t$ by making many function calls for different values of $\vec b$. In \sartre, we tabulate the DGLAP evolution of the gluon density in $\xpom$ and $\mu^2$ which speeds up the cuhre integration by two orders of magnitude, but this interpolation also introduces noise in the integration, which sometimes lead to numerical glitches. To make the comparison between the two methods as fair as possible, we performed the calculation twice, once with the full DGLAP evolution in each point in order to compare the results of the integrals, and once with the DGLAP lookup table in order to compare the speeds. The precision of the FFT was not affected by choice of DGLAP evaluation. 
This ensures that we get a valid test of the precision of the integration method, as well as a comparison of timings which are relevant for the user. All models use the same nuclear configuration for the calculation.
In table \ref{tab:table1}, we see that we have an agreement of $1-5\%$ between the two numerical integrations in all kinematic points considered. We use the $t$ dependence in $\xpom(t)$ as described in eq.\eqref{eq:slab}. Without it, the difference between the calculations climb to $\sim10\%$ for the larger $t$-values.

In table \ref{tab:times} we report on the timings to produce the calculations in tab.\ref{tab:table1}. We present an average over six calculations using the same machine. We have made sure that all calculations run on a single core, since all methods can be paralellized to an arbitrary degree. We present the Cuhre result in seconds and the DFT and FFT results in ms. We see that the speed-up using a DFT gives an increase by 2,000-7,100 times. By comparison the tabulated FFT increase of 720-3,500 times looks modest, but this method calculates all $N_t=N_b/2$ points at once, so we compare with a calculation of 5 points with a calculation of 128 and 256 points respectively. An increase of 2,000 reduces one week of running to 5 minutes.  

\subsection{Predictions for the Electron-Ion Collider and the Large Hadron-electron Collider}
\label{sec:predictions}
For this paper we have created several sets of new tables in $e$Pb scattering, with $N_{Q^2}\times N_W=30\times 30$ bins. We have considered three regions in $t$. Region 1 covers the first two coherent dips with $|t|<0.04$. Region 2 covers the region in which  fluctuations at the scale of nucleons are resolved, which is $0.04\lesssim |t|\lesssim0.4~{\rm GeV}^2$, and region 3 covers the region that is sensitive to hotspot fluctuations $0.4\lesssim|t|\lesssim2.5$. The hotspot model does not describe HERA measurements for $|t|>2.5~{\rm GeV}^2$ \cite{Kumar:2024kns}. We use the DFT approach in all three regions which enables us to choose our binning in $t$. We produce three sets of tables that overlap slightly around the region edges, and then we merge the tables into a single set of tables for the entire $t$-spectrum for $0<|t|<2.5~{\rm GeV}^2$.

In region 1, we set $b_{\rm max}=3R_A$, where $R_A$ is the nuclear radius. It should be noted that $b_{\rm max}\gtrsim 1.8R_A$ constitute zero-padding in the DFT integrand. This can be used to increase the resolution in the reciprocal grid, but comes at the cost of a decreased $\vec b$-resolution. In region 1 we have to find a balance between good $t$-resolution and being able to resolve the Woods-Saxon skin depth well enough not to affect the resulting cross-sections. We choose $N_b=128$ and average over 200 configurations, which ensures good convergence in this region, and we use $N_t=50$ user specified linearly spaced $t$-bins.

In region 2 we reduce $b_{\rm max}$ to $b_{\rm max}=2.5R_A$ and increase the number of configurations to 500. We also increase $N_b$ to 256 and set $N_t=22$. 

For region 3, we set $b_{\rm max}=2.2R_A$, average over 600 configurations, keep $N_b=256$ and calculate the amplitude moments at $N_t=18$ points. This gives a $b$-resolution of $\delta b\approx0.11$fm which is enough to resolve a hotspot width of $\sqrt{B_q}\approx0.22$fm. 

It should be noted that we could have opted for the FFT approach as well. In that case, the binning in $t$ is given by $\delta t = 2\delta\Delta\sqrt{|t|}$, where $\delta\Delta=2\pi/(N_b\delta b)=\pi/b_{\rm max}$, i.e. the $t$-binning of the reciprocal grid depends \emph{only} on $b_{\rm max}$, and not on $N_b$ or $\delta b$. When using this method, one needs therefore to be careful with setting the parameters such that one gets enough precision both in $\delta t$ and in $\delta b$. For example, one may need to both resolve the coherent peak well in $t$, for which the precision in $b$ needs to resolve the Woods-Saxon skin depth of $\si  0.5$fm. The reciprocal grid runs from $|t|=(\pi/b_{\rm max})^2$, to $|t|=(N_b\pi/2b_{\rm max})^2$. For the region 3 parameters above ($b_{\rm max}=2.2R_A\approx 14.6$fm, $N_b=256$), this would correspond to $N_t=128$ bins, distributed in the region $0.0018\leq|t|\leq29.6$~GeV$^2$. These points are denser towards the lower part of the range, and out of these, 37 points are withing $|t|<2.5$~GeV$^2$. The speedup from using an FFT rather than an arbitrary $t$-grid DFT is modest, the FFT scales as $\mathcal{O}(N_b\log_2N_b)$ and the DFT as $\mathcal{O}(N_t\times N_b)$, and most $t$-bins are not used in the FFT case. If one use the same 37 points in a DFT the computation times are comparable.

\begin{figure}
    \centering
    \includegraphics[width=0.45\linewidth]{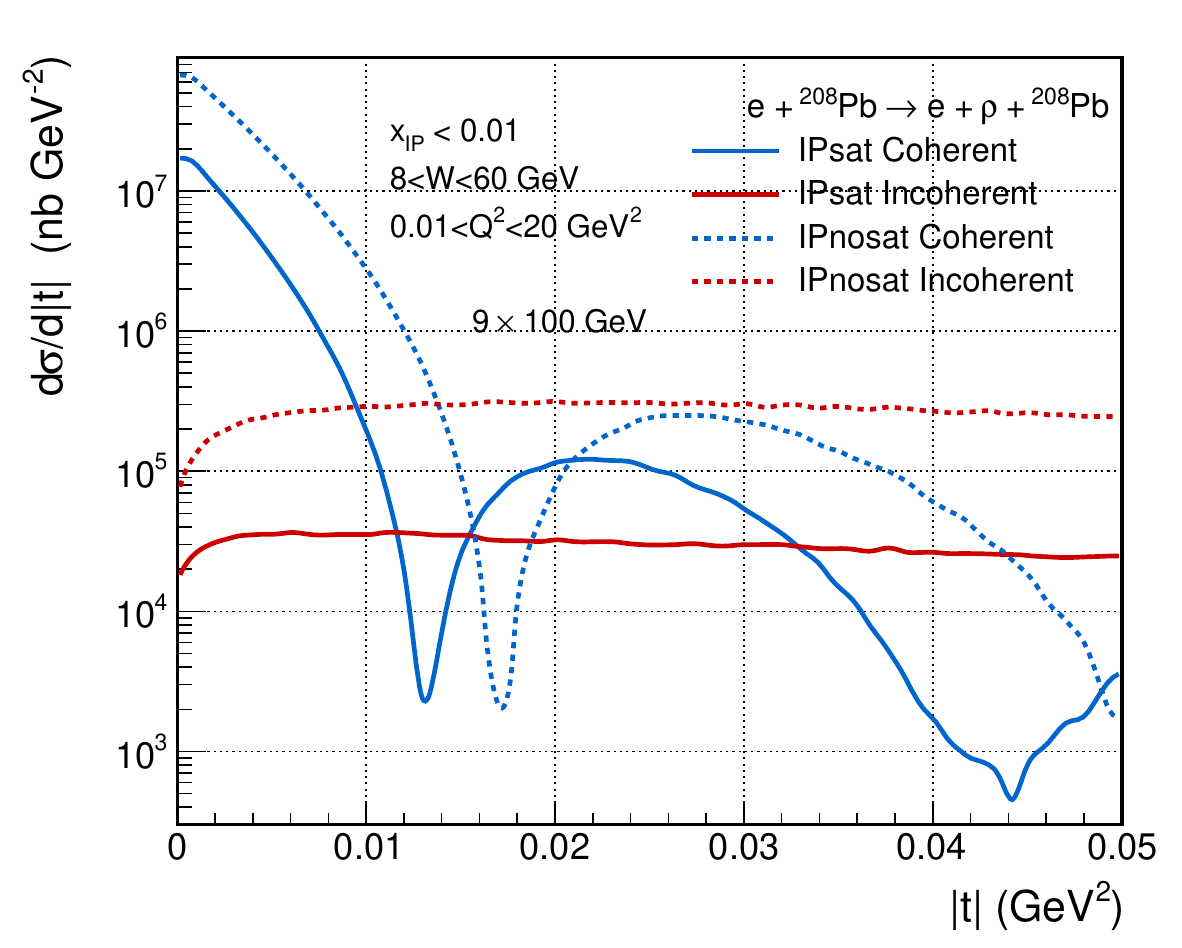}
    \includegraphics[width=0.45\linewidth]{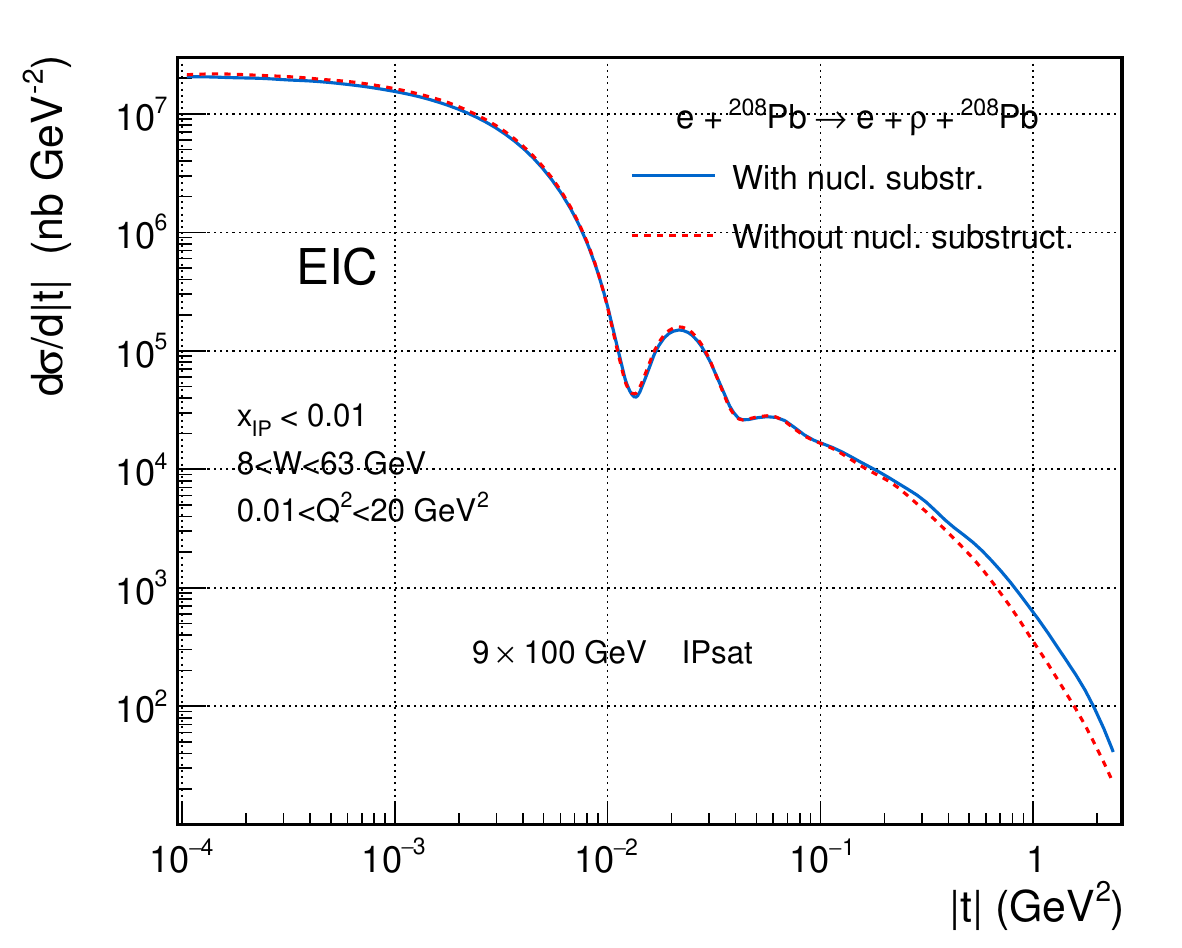}
    \caption{Predictions for $\rho$ production at the EIC using \sartre. Left side: Comparison between IPsat and IPnosat for small $|t|$. Right side: comparison between IPsat with and without nucleon hotspot substructure.}
    \label{fig:EIC}
\end{figure}
In figure \ref{fig:EIC} we show the results of 10M events generated using tables produced with the new method. The left-side of the fig. \ref{fig:EIC} show exclusive $\rho$-production with both IPsat and IPnosat for small $|t|$ at $\sqrt{s}=60~{\rm GeV}$, which is the planned initial energy at the Electron-Ion Collider (EIC) \cite{EICtalkDIS}. Since $\rho$ has a large dipole size, it is more sensitive to non-linear effects than heavier vector mesons, which is clearly visible in the figure. However, for the same reason, $\rho$-meson production is not a good probe for detailed nuclear structure, such as nucleon substructure, as can be seen in the right side of fig.\ref{fig:EIC}. Here, we show $\rho$ production with and without subnucleon hotspot structure, and the two curves are close to each other. This graph still demonstrates \sartre's capabilities to generate events in a large kinematic range for the EIC.

\begin{figure}
    \centering
    \includegraphics[width=0.45\linewidth]{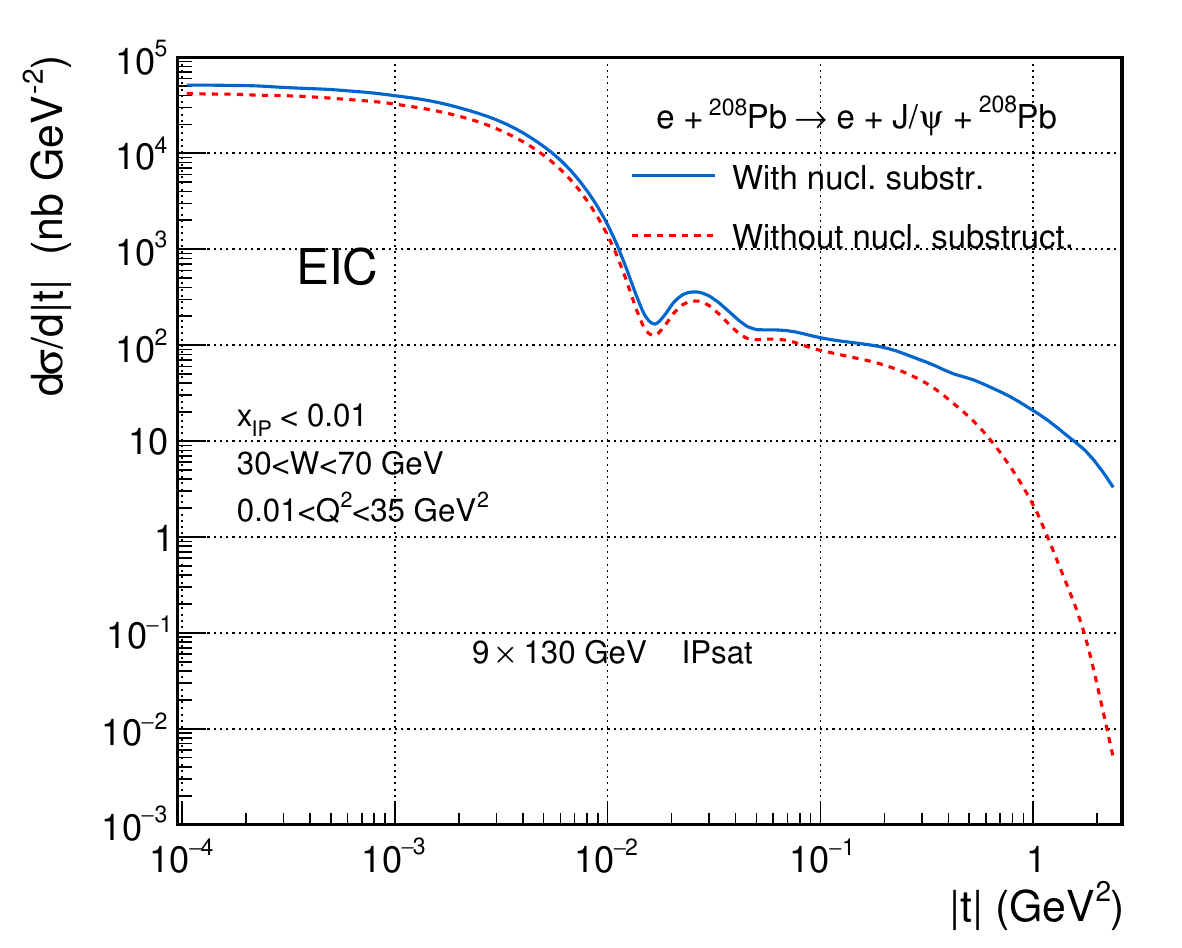}
    \includegraphics[width=0.45\linewidth]{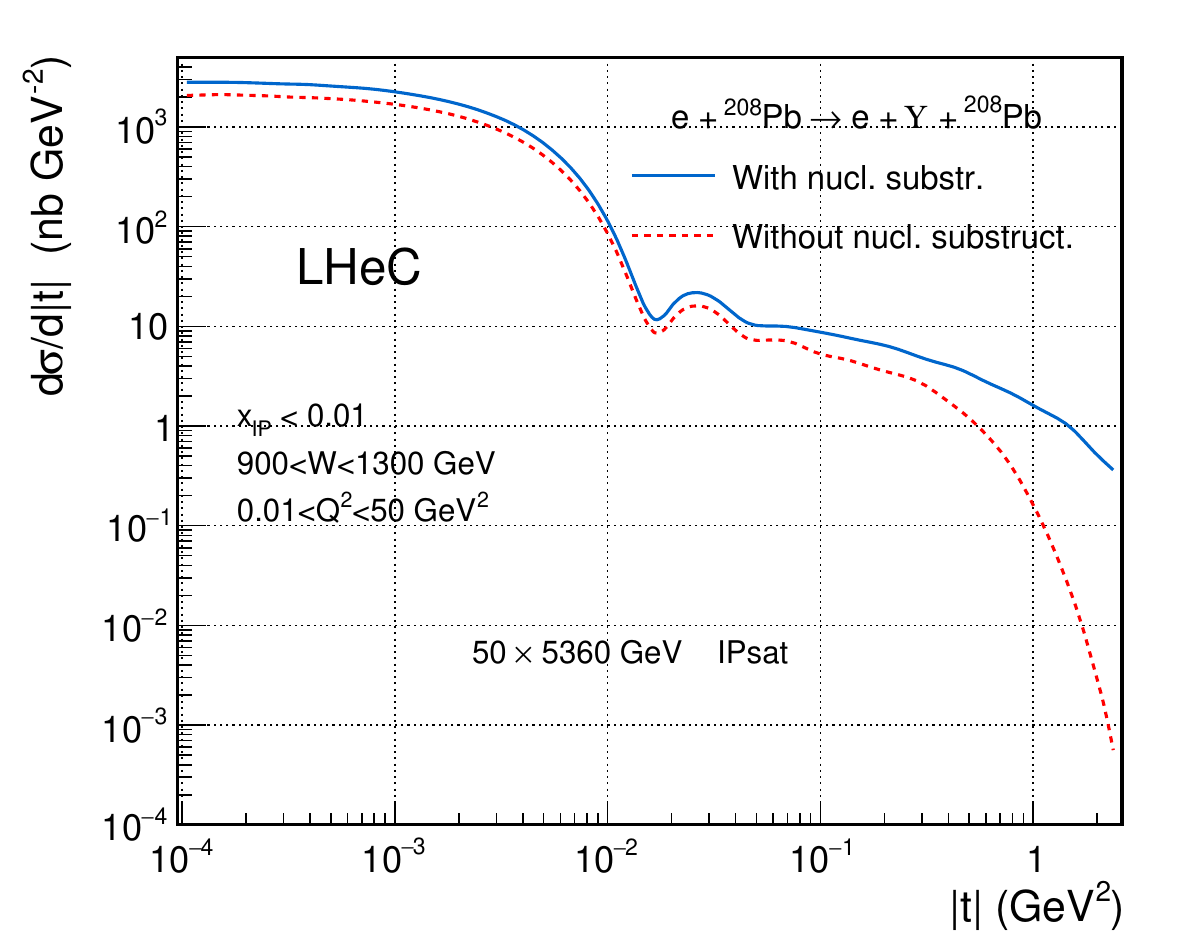}
    \caption{Predictions for heavy meson production at the EIC and the LHeC in the IPsat model with and without nucleon hotspot substructure. Left side: $J/\psi$ production at the EIC. Right side: $\Upsilon$ production at the LHeC.}
    \label{fig:LHeC}
\end{figure}

For observing nuclear substructure one needs heavy vector meson production. On the right-hand side of fig.~\ref{fig:LHeC} we show $J/\psi$ production in the IPsat model with and without nucleon substructure. We show $J/\psi$ production at a later planned energy setting with $\sqrt{s}=68.4~{\rm GeV}$. Here we see a clear separation between the predictions with nucleon substructure and without, at large $|t|$.

The ultimate probe for substructure, due to its very small size, would be the $\Upsilon$ meson, as shown in \cite{Demirci:2022wuy}. To produce $\Upsilon$ mesons at $\xpom<0.01$ one needs $W\gtrsim 90$~GeV for moderate $Q^2$, which is not achievable at the EIC. At the LHeC ~\cite{LHeC:2020van}, on the other hand, energies of $\sqrt{s}=1,035~{\rm GeV}$ may become available. At those energies, $\Upsilon$-production becomes feasible, which due to its very small size is an excellent probe for nuclear structure. We show the prediction from \sartre~ for this process at the LHeC, and we see that there is a pronounced difference if we use Gaussian nucleons or nucleons with hotspots in our event generator.

\section{Summary and Outlook}
\sartre~ is an event generator for exclusive diffraction at small $x$ in $ep$, $e$A, as well as ultra-peripheral $pp$, $p$A and AA collisions. \sartre~ precomputes the first and second moments of the dipole amplitude and stores them in lookup tables, in a grid in $Q^2$, $W^2$, and $t$. This enables efficient event generation of both coherent and incoherent diffraction. However, these lookup tables has previously taken CPU-years of production time for each combination of initial state target nucleus and final state vector meson (or DVCS photon). Further, these tables often contained numerical glitches that caused spikes in event generation for some points. In a previous study, we reduced the production time of lookup tables by one order of magnitude, as well as smoothening the tables, by utilizing machine learning techniques ~\cite{Singh:2023yvj}.

In this paper we have presented an improved calculation exhibiting 3-4 orders of magnitude improvement in efficiency, reducing each production to $\sim1$ CPU-day, which with parallelization becomes a feasible calculation on a single computer, rather than necessitating the use of farm computing.

This has been achieved with a combination of techniques. The largest improvement comes from turning the impact-parameter $\vec b$ integral into a discreet Fourier transform (DFT) or a fast Fourier transform (FFT). This takes advantage of the fact that $\Delta=\sqrt{-t}$ is the Fourier reciprocal of $\vec b$. Using the projection-slice theorem further turns it into a one-dimensional Fourier transform. This enables us to only calculate the Fourier transform at the points in $t$ that we are interested in, instead of performing a full two-dimensional numerical integration for each point in $t$. We also further optimize the numerical calculation by precomputing all factors that are in common for each point in $\vec b$. We have implemented both DFT, enabling arbitrary $t$-binning and FFT, which calculates $N_t=N_b/2$ $t$-values on the reciprocal grid. 

The overall improvement in efficiency finally enables \sartre~to simulate all possible physics at small $x$ exclusive diffraction at the EIC and the LHeC. We have demonstrated this by showcasing completely new tables for $e$Pb, scattering and showing what physics it enables. For light vector mesons we have shown the impact of non-linear effects already at the initial energies of the EIC of $9\times 100$GeV, and the capabilities of heavy vector mesons to probe the fine details of nuclear structure at the EIC, with $J/\psi$ mesons, or the LHeC where $\Upsilon$ mesons become kinematically available at small $\xpom$. 

\section*{Acknowledgements}
The authors acknowledge the support from the physics department of IIT Delhi. This research was supported by Core Research Grant (CRG) support CRG/2022/002507 from Anusandhan National Research Foundation (ANRF), Department of Science and Technology, Government of India. 
\bibliographystyle{unsrt}
\bibliography{Saturationandfluctuations}

\end{document}